# A Family of Likelihood Ascent Search Multiuser Detectors: an Upper Bound of Bit Error Rate and a Lower Bound of Asymptotic Multiuser Efficiency

Yi Sun[1]

*Abstract* – **In this paper, the bit error performance of a family of likelihood ascent search (LAS) multiuser detectors is analyzed. An upper bound on the BER of any LAS detector is obtained by bounding the fixed point region with the worst initial detector. The concept of indecomposable errors developed by Verdú is applied to tighten the upper bound. In a special instance, the upper bound is reduced to that for all the local maximum likelihood detectors. The upper bound is comparable with that of the optimum detector obtained by Verdú. A lower bound on the asymptotic multiuser efficiency (AME) is then obtained. It is shown that there are nontrivial CDMA channels such that a LAS detector can achieve unit AME regardless of user number. The AME lower bound provides a means for further seeking a good set of spreading sequences and power distribution for spectral and power efficient CDMA.**

## I. INTRODUCTION

The global maximum likelihood (GML) multiuser detector[2] is optimum in terms of minimization of error probability in joint detection of multiuser symbols in a CDMA channel [1]-[3]. As a problem searching the closest point in lattices [4], the GML multiuser detection is NP-hard in general. To trade complexity with error performance, numerous linear, nonlinear and hybrid linear-nonlinear low-complexity suboptimal detectors have been developed in the past over two decades. For the linear detectors, the MF, decorrelator, and MMSE are most famous and their performance is thoroughly analyzed in literature. There are a large number of nonlinear iterative detectors [5]-[14]. Though having a complexity much lower than NP-hard and performing comparably with or better than the famous linear detectors, they still perform far from optimal and most of them have a per-bit complexity higher than linear in user number $K$, thus prohibited in fairly large systems. For instances, with fixed spreading sequences the per-bit complexity is $O(K^2)$ for PDA [9], $O(K^{2.5})$ for SDR [10], and $O(K^2)$ for SDP [12], and in contrast only $1.5K$ for the MMSE-DF [1]. The average complexity of the sphere decoding [15][16] is cubic or higher in $K$ for a moderate SNR [17].

Current research has focused on practical multiuser detectors with performance close to optimal. To this end, the belief propagation (BP) is recently applied to multiuser detection [18]-[20]. The BP iteratively increases *a posterior* probability of tentatively estimated bit vector and is shown to converge to the GML in large random spreading CDMA. To reduce the original BP complexity growing exponentially in $K$, Gaussian approximation is necessarily employed [18]. A nonlinear MMSE detector with Gaussian approximation is similarly developed in [21].

Here we introduce a family of likelihood ascent search (LAS) detectors with several advantages for multiuser detection. One of the main advantages is its low per-bit complexity – less than $0.5K$ demonstrated in various simulations. Another is that a LAS detector can approach BER of the GML detector in large random spreading CDMA [39] when $K$ is greater than 500 and the channel load is as high as 1 bit/s/Hz. For $K < 500$, a scheme of bit multiplexing and extending is proposed to construct quasi-large random spreading CDMA without incurring either increase of frequency bandwidth and/or transmission power or decrease of transmission rate. Hence, the LAS detector with a practical complexity can approach the GML BER and single-user performance (in high SNR regime) in any practical system [39].

---

[1] Yi Sun is with the Department of Electrical Engineering at the City College of City University of New York, New York, NY 10031. E-mail: ysun@ee.ccny.cuny.edu. This paper was presented in part at the 41st Annual Conference on Information Science and Systems, Baltimore, Maryland, March 14-16, 2007.
[2] To discriminate the local maximum likelihood detector in this paper, the joint optimum detector is termed the GML detector.

The family of LAS detectors was originally developed for digital image restoration [23]-[28]. Due to huge dimension of a digital image (a 512 by 512 image has a dimension of 262,144 pixels), iterative algorithms for image restoration must have a complexity not higher than linear. With the advantages of linear complexity and suitability for parallel implementation on hardware for fast computation, the HNN based algorithms are developed in [29]. To overcome the instability inherited to the original HNN, the modified HNN (MHNN) algorithms are proposed in [30][23][24] and are further developed into the family of MHNN algorithms [25]-[28]. When applied to multiuser detection, these algorithms turn out to perform likelihood ascent search and thereafter called the LAS detectors [31][32]. The previous studies [23]-[28][30]-[32] have shown that the LAS detectors are self-contained to converge to a fixed point in a finite number of steps. A LAS detector always searches out a sequence of vectors with monotonic likelihood ascent.

The wide-sense sequential LAS (WSLAS) detectors, including the sequential LAS (SLAS) detector, are the best in the family that achieve local maximum likelihood (LML) in a neighborhood of size one. It is not surprising due to its simplicity that the SLAS detector is also developed with different motivations and named differently in literature [5][33][34]. Unlike the GML detector, the LML detectors are not unique with greatly different performances. To improve the SLAS performance, the eliminating-highest-error (EHE) and the fastest-metric-descent (FMD) algorithms are developed in [24][25][31]. The algorithms with neighborhood size one in [35][36] are identical to the FMD algorithm [31]. All the gradient guided search algorithms in [35] and the WSLAS detectors fall into the class of LML detectors with *any* neighborhood size [37].

Pioneering the area of multiuser detection, Verdú found an upper bound of GML BER [1][2], which indicates the vast performance gap between the GML and the conventional MF detectors. He also proposed the asymptotic multiuser efficiency (AME) as a performance measure in high SNR regime [1][3]. A multiuser detector can achieve the single-user bound in high SNR regime, as if there was no interferer, if and only if its AME equals one. There exist CDMA channels where the GML detector can achieve unit AME [1][38]; in contrast, the famous linear detectors perform far worse than the single-user bound.

Although a large number of nonlinear iterative detectors have been developed in literature, their performance analysis is difficult, i.e. no BER upper bound as well as AME has been found. In the past over two decades, it has not been evidenced yet that AME can be successfully applied to design of CDMA systems with high efficiency of spectrum and power. However, this does not mean that AME is useless in practical CDMA system design; instead, this is due to that most nonlinear iterative detectors are heuristically developed with unknown decision regions in general. Moreover, for long the most important ingredients of CDMA systems – power control, spreading sequences, and multiuser detector – have been separately designed, thus hindering development of high spectral and power efficient CDMA. This is due also to that analytical performance as a function of spreading sequences and power distribution is intractable to obtain for most nonlinear detectors.

In this paper, we obtain a BER upper bound for the family of LAS detectors, thanks to that the fixed point region of any LAS detector can be obtained. The bound is obtained by bounding the union of all fixed point regions and thus is the tightest for the worst LAS detector with any initial. The upper bound is comparable with the GML bound by Verdú [1]. The performance difference between the LAS and GML detectors is clearly indicated by the upper bound, due to the shorter distance of the true signal to the LAS fixed point region of an error signal. A lower bound of AME for any LAS detector is also obtained. In a special instance, the bounds become those for the LML detectors. It is found that there are nontrivial CDMA channels where regardless of user number the LAS detectors can achieve unit AME and thus achieve the single-user performance in high SNR regime. In [39], by analysis of the bounds it is proved that the AME of the WSLAS detectors converges almost surely to one in the limit of large random spreading CDMA with channel load less than ½ − 1/(4ln2) bits/s/Hz. It is expected that by means of the LAS BER and AME bounds, more sets of spreading sequences and power distributions shall be found to achieve high efficiency of spectrum and power in CDMA. In return, the findings in this paper and [39] suggest that AME be useful in design of spectral and power efficient CDMA.

The rest of the paper is organized as follows. In Section II, the LAS detectors are developed with study of fixed point region. In Section III, the BER and AME bounds are analyzed. The BER bound is proved in Section IV. Conclusions are made in Section V and all other proofs are presented in the Appendix.

## II. THE LAS DETECTORS

### A. CDMA signal

The CDMA channel model and notations in this paper are the same as in [1]. Consider a $K$-user bit-synchronous Gaussian CDMA channel. The chip matched filter (MF) at the receiver outputs a real vector

$$\mathbf{r} = \mathbf{SAb} + \mathbf{m}. \qquad (1)$$

$\mathbf{S} = (\mathbf{s}_1, ..., \mathbf{s}_K) \in \mathbb{R}^{N \times K}$ where $\mathbf{s}_k$ is the $k$th user's spreading sequence with unit length $\|\mathbf{s}_k\| = 1$ and spreading factor $N$. $\mathbf{A} = \text{diag}(A_1, ..., A_K)$ where $A_k$ is the received signal amplitude. $\mathbf{b} \in \{-1, 1\}^K$ is the vector of transmitted bits of $K$ users independently equiprobably taking on $\pm 1$'s. $\mathbf{m} \sim N(\mathbf{0}, \sigma^2 \mathbf{I})$ is a white Gaussian noise vector. The MF bank $\mathbf{S}$ outputs a sufficient statistic

$$\mathbf{y} = \mathbf{S}^T \mathbf{r} = \mathbf{RAb} + \mathbf{n}. \qquad (2)$$

where $\mathbf{R} = \mathbf{S}^T \mathbf{S}$ is the sequence crosscorrelation matrix and $\mathbf{n} = \mathbf{S}^T \mathbf{m} \sim N(\mathbf{0}, \sigma^2 \mathbf{R})$. A likelihood function is $f(\mathbf{y}|\mathbf{b}) = -\tfrac{1}{2}\mathbf{b}^T \mathbf{H} \mathbf{b} + \mathbf{b}^T \mathbf{A} \mathbf{y}$ with $\mathbf{H} = \mathbf{ARA}$ where $H_{kk} = A_k^2$ since $R_{kk} = 1$ for any $k$. The GML detector $\mathbf{b}^{\text{GML}}$ achieving the maximum likelihood in $\{-1,1\}^K$ is optimum in terms of minimizing error probability $P_e(\mathbf{b}^\phi) \equiv \Pr(\mathbf{b}^\phi \neq \mathbf{b})$ among all detectors $\phi$.

*B. Criteria in design*

Given $\mathbf{b}(n)$ at search step $n$, a LAS detector updates a number of bits and obtains a new bit vector $\mathbf{b}(n+1)$ until reaching a fixed point. The bits scheduled to update in a step are called update candidates. Updating a bit is to check the flip condition of the bit but may or may not result in a bit flip.

Suppose $L(n) \subseteq \{1, ..., K\}$ is the index set of update candidates. The likelihood gradient evaluated at $\mathbf{b}(n)$ equals

$$\mathbf{g}(n) = -\mathbf{H}\mathbf{b}(n) + \mathbf{A}\mathbf{y} \qquad (3)$$

which multiplied by $\mathbf{A}^{-1}$ is simply the difference between the MF bank output and the CDMA signal reconstructed by $\mathbf{b}(n)$. If the bits in $L_p(n) \subseteq L(n)$ are flipped, then $\mathbf{b}(n+1) = \mathbf{b}(n) - 2\sum_{i \in L_p(n)} b_i(n)\mathbf{e}_i$ where the $i$th element of $\mathbf{e}_i$ equals one and others are zero. The likelihood gradient can be efficiently updated by

$$\mathbf{g}(n+1) = \mathbf{g}(n) + 2\sum_{i \in L_p(n)} b_i(n)\mathbf{H}_i . \qquad (4)$$

The likelihood change $\Delta f(n) = f[\mathbf{b}(n+1)] - f[\mathbf{b}(n)]$ can be calculated in terms of $\mathbf{g}(n)$ as

$$\Delta f(n) = \Delta \mathbf{b}^T(n)[\mathbf{g}(n) + \tfrac{1}{2}\mathbf{z}(n)] \qquad (5)$$

with $\Delta \mathbf{b}(n) = \mathbf{b}(n+1) - \mathbf{b}(n)$ and $\mathbf{z}(n) = -\mathbf{H}\Delta \mathbf{b}(n)$.

The LAS detector is developed based on the criteria that (i) the updating rule is computationally efficient; (ii) the new vector $\mathbf{b}(n+1) \neq \mathbf{b}(n)$ must have higher likelihood than $\mathbf{b}(n)$; and (iii) under the same framework of updating rule, $\mathbf{b}(n+1)$ has the highest likelihood. In what follows, we briefly discuss how a LAS detector satisfies the criteria.

After the flip of bits in $L_p(n)$, the likelihood gradient can be efficiently updated by (4), which requires $|L_p(n)|K$ additions. Employing $\mathbf{g}(n)$ in the updating rule makes computation simple and efficient. $\mathbf{g}(n)$ indicates the direction along which likelihood increases. However, the bit vectors belong to a finite set and $\mathbf{g}(n)$ can only roughly point at the direction along which the next higher-likelihood vector is located. The updating rule $\mathbf{b}(n+1) = \text{sgn}[\mathbf{g}(n)]$ of the original HNN may enter a limit cycle rather than ensure likelihood ascent. To understand this, we note that (5) can be rewritten as

$$\Delta f(n) = \sum_{k \in L(n)} \Delta b_k(n) \left[ g_k(n) + \frac{1}{2} z_k(n) \right]. \qquad (6)$$

Clearly, the original HNN cannot guarantee $\Delta b_k(n)[g_k(n) + \tfrac{1}{2}z_k(n)] \geq 0$ for all $k \in L_p(n) \subseteq L(n)$. If the sum for all $\Delta b_k(n)[g_k(n) + \tfrac{1}{2}z_k(n)] > 0$ is not greater than the negative sum for all $\Delta b_k(n)[g_k(n) + \tfrac{1}{2}z_k(n)] \leq 0$, then $\Delta f(n) \leq 0$, implying an unstable updating rule. To satisfy (ii), only those bits that satisfy $\Delta b_k(n)[g_k(n) + \tfrac{1}{2}z_k(n)] > 0$ should be updated. However, $z_k(n)$ depends on all the bits in $L_p(n)$, which are unknown before all the candidates are updated. To avoid the paradox, the LAS detector considers the worst case when all candidates are flipped (i.e. $L_p(n) = L(n)$). Then a threshold for each candidate can be properly set up in the updating rule.

*C. The LAS detector*

The following generalized LAS detector that is applicable to all possible sequences of sets of candidates defines the family of LAS detectors.

*LAS detector*: Given $L(n) \subseteq \{1, ..., K\}$ for all $n \geq 0$ and an initial vector $\mathbf{b}(0) \in \{-1, 1\}^K$. At step $n$ all the bits for $k \in L(n)$ are updated by

$$b_k(n+1) = \begin{cases} +1, & \text{if } b_k(n) = -1, \text{ and } g_k(n) > t_k(n), \\ -1, & \text{if } b_k(n) = +1, \text{ and } g_k(n) < -t_k(n), \\ b_k(n), & \text{otherwise,} \end{cases} \quad (7)$$

where the $k$th threshold is

$$t_k(n) = \sum_{j \in L(n)} |H_{kj}|, \quad (8)$$

all the bits for $k \notin L(n)$ remain unchanged $b_k(n+1) = b_k(n)$, and then $\mathbf{g}(n+1)$ is updated by (4) in which $L_p(n)$ is the index set of flipped bits in (7). $\mathbf{b}^*$ is the finally demodulated vector if $\mathbf{b}(n) = \mathbf{b}^*, \forall n \geq n^*$ with some $n^* \geq 0$. □

The thresholds in the LAS detector are not arbitrary as will be shown in the next section.

In practice, $L(n)$ can be scheduled such that all the bits are finally periodically updated. If no bit is flipped in a period, the LAS detector has reached a fixed point $\mathbf{b}^*$ and shall terminate. Throughout, assume that before reaching a fixed point, every bit is updated once more without flip. We consider only deterministic $L(n)$ though some results are applicable to random $L(n)$. The LAS detector is time-invariant if $L(n_1) = L(n_2)$ for all $\mathbf{b}(n_1) = \mathbf{b}(n_2)$, $n_1 \neq n_2$.

Specifying a sequence of $L(n)$ for $n \geq 0$, one determines a particular LAS detector. One of the simplest sequences is to update one bit in each step, which produces a SLAS detector with the lowest threshold $t_k = A_k^2$. The SLAS detector can update the bits in a circular or random order. All the SLAS detectors belong to the larger set of WSLAS detectors that have $|L(n)| = 1, \forall n \geq n'$ with some $n' \geq 0$. Another simplest sequence is $L(n) = \{1, ..., K\}, \forall n$, which yields the parallel LAS (PLAS) detector with the highest threshold $t_k = \sum_{j=1}^{K} |H_{kj}|$. The PLAS and the SLAS detectors are special instances of the group-parallel LAS (GPLAS) detectors that update bits group by group. If $\varsigma$ is a collection of subsets partitioning $\{1, ..., K\}$, a GPLAS detector has $L(n) \in \varsigma$ with $t_k = \sum_{j \in L} |H_{kj}|, \forall k \in L \in \varsigma$.

The initial $\mathbf{b}(0)$ can be any detector. However, an initial with lower error probability can make the LAS detector converge faster to a fixed point of lower error probability. To reduce dependency on the initial computational cost, the MF or random vector can be employed. In the rest analysis, the initial is assumed arbitrary unless specified.

The updating rule (7) with the gradient update (4) being the core cost of computation is the key to make the LAS detector linearly complex. The total number of bit flips from $\mathbf{b}(0)$ to $\mathbf{b}^*$ equals $M = \sum_{n=0}^{n^*} |L_p(n)|$, which depends on $L(n)$, $\mathbf{b}(0)$, and $\mathbf{y}$ and is random. We define the bit flip rate (BFR) as the average number of flips per bit $c = E(M)/K$. The complexity can be defined as the average number of additions per bit. Each bit flip results in $K$ additions in (4). The complexity equals $E(KM)/K = E(M) = cK$ linear in $K$. The BFR $c$ is less than 0.81 as observed in all simulations in various conditions, part of which is reported in [39]. With implementation on hardware suitable for vector additions, the computation time is further reduced by a factor of $K$.

Equation (7) is elaborately written in the form for convenience of digital hardware implementation. Though (7) can be also rewritten without use of **g**, there are good reasons to use. The efficient updating of **g** in (4) is the key to reduce redundant computations, and moreover it can be implemented on hardware suitable for fast parallel computation. In addition, searching next vector along **g**($n$) with likelihood ascent is the motivation, which identifies the LAS detectors from others motivated by different criteria [5][7][33][34].

*D. Monotonic likelihood ascent*

The following theorem indicates that a LAS detector monotonically increases likelihood and reduces error probability. Unless the initial is a fixed point with probability one, the LAS detector reduces the initial error probability. The proof of (i) can be obtained in the similar way in [26] where $b_i$ takes on 0, 1, and (ii) and (iii) are further obtained due to Lemma 5 in the Appendix. Throughout, an LML point **b**$^{\text{LML}}$ satisfies $f(\mathbf{y} \mid \mathbf{b}^{\text{LML}}) \geq f(\mathbf{y} \mid \mathbf{a})$ for all **a** that differ from **b**$^{\text{LML}}$ by at most *one* bit unless otherwise specified.

*Theorem 1*: Consider a LAS detector with $L(n)$, $n \geq 0$ that generates **b**($n$), $n \geq 0$. (i) For any **y**, $f[\mathbf{y} \mid \mathbf{b}(n+1)] \geq f[\mathbf{y} \mid \mathbf{b}(n)]$, $\forall n \geq 0$ with equality iff **b**($n+1$) = **b**($n$); (ii) $P_e[\mathbf{b}(n+1)] \leq P_e[\mathbf{b}(n)]$ with equality iff $\Pr[\mathbf{b}(n+1) = \mathbf{b}(n)] = 1$; (iii) $P_e(\mathbf{b}^*) \leq P_e[\mathbf{b}(0)]$ with equality iff **b**(0) is a fixed point with probability one, and for the WSLAS detector with equality iff **b**(0) is an LML with probability one. □

*E. Fixed-point region*

Given $L(n)$ for $n \geq 0$, in the **y** space there is a region of **b** associated with the initial detector **b**(0) where the LAS detector converges from **b**(0) to **b**. To analyze the effect of a particular sequence $L(n)$ on error performance, we consider the union of such regions associated with all initial detectors, which is called the *fixed-point region* $V^{\text{LAS}}(\mathbf{b})$ ≡ {**y** ∈ $\Re^K$ | there is **b**(0) s. t. the LAS detector converges to **b**}.

*Proposition 1*: For any **y**, denote by $\Lambda(\mathbf{y},\mathbf{b}) = \{\mathbf{b}(0) \in \{-1, 1\}^K \mid \text{LAS: } \mathbf{y}, \mathbf{b}(0) \to \mathbf{b}\}$ the set of initial vectors from which the LAS detector converges to **b**. Let $t_k^* = \max_{\mathbf{b}(0) \in \Lambda(\mathbf{y},\mathbf{b})} \min_{n \geq n_k^*[\mathbf{b}(0)], k \in L(n)} t_k(n)$ where $n_k^*[\mathbf{b}(0)]$ denotes the last flip step of the $k$th bit with the initial **b**(0). Then

$$V^{\text{LAS}}(\mathbf{b}) = \{\mathbf{y} \in \Re^K \mid \mathbf{b} \otimes (\mathbf{Ay} - \mathbf{Hb}) \geq -\mathbf{t}^*\} \tag{9}$$

where $\mathbf{t}^* = (t_1^*,...,t_K^*)^T$ and multiplication $\otimes$ and inequality $\geq$ are element-wise. □

The thresholds of the SLAS, PLAS, and GPLAS detectors all are time-invariant and therefore $t_k^* = t_k$ where $t_k$ is given accordingly in Subsection C. A WSLAS detector has the same fixed-point region of the SLAS detector as

$$V^{\text{WSLAS}}(\mathbf{b}) = \{\mathbf{y} \in \Re^K \mid \mathbf{b} \otimes (\mathbf{y} - (\mathbf{R} - \mathbf{I})\mathbf{Ab}) \geq \mathbf{0}\} \tag{10}$$

which is also the LML region $V^{\text{LML}}(\mathbf{b}) \equiv \{\mathbf{y} \in \Re^K \mid \mathbf{b} \text{ is an LML point}\}$. Hence, all WSLAS detectors are also LML detectors.

The fixed point region (9) shall be applied to obtain the BER upper bound in the next section. Before proceeding, by Proposition 1 we qualitatively analyze characteristics of the fixed point region and the relationship between the error performance and the thresholds. First, any bit vector **b** can be a fixed point with a nonzero probability. Second, given **y**, the GML decision $\mathbf{b}^{GML}(\mathbf{y})$ is unique and the GML decision regions $V^{GML}(\mathbf{b})$ for different **b**'s do not overlap. This is also true for other detectors such as linear detectors. In contrast, a LAS detector may have a set $\Psi^{LAS}(\mathbf{y})$ of fixed points, and the fixed-point regions $V^{LAS}(\mathbf{b})$ for different **b**'s may overlap. Similarly, there may be a set $\Psi^{LML}(\mathbf{y})$ of LML points, and LML regions for different bit vectors may overlap. In the overlapped fixed-point region, one of the fixed points is taken as the demodulated vector depending on the initial vector and $L(n)$. Third, as the thresholds increase, the fixed-point region expands, the overlapped region expands, and the number of fixed points in the expanded region increases. Since the increased fixed points have lower likelihood, the error probability increases. On the other hand, increasing the number of low-likelihood points makes it easy to reach a fixed point, thus decreasing computational complexity.

The following duality is obtained from the proposition: (i) For any **b**, $V^{GML}(\mathbf{b}) \subseteq V^{WSLAS}(\mathbf{b}) = V^{LML}(\mathbf{b}) \subseteq V^{LAS}(\mathbf{b}) \subseteq V^{PLAS}(\mathbf{b})$. All the equalities in $\subseteq$ hold iff **R** = **I** where all the LAS detectors as well as the GML detector are collapsed to the $K$ parallel single-user MF detectors; (ii) for any $\mathbf{y} \in \Re^K$, $\{\mathbf{b}^{GML}(\mathbf{y})\} \subseteq \Psi^{WSLAS}(\mathbf{y}) = \Psi^{LML}(\mathbf{y}) \subseteq \Psi^{LAS}(\mathbf{y}) \subseteq \Psi^{PLAS}(\mathbf{y})$. The equalities in (ii) can be true for some **y** even when **R** ≠ **I**. For example, if $|A_k y_k| > \sum_{j=1}^{K} |H_{kj}| + t_k^* - 2H_{kk}$ for all $k$, then the GML point is the only fixed point, $\Psi^{LAS}(\mathbf{y}) = \{\mathbf{b}^{GML}(\mathbf{y})\}$ and $b_k^{GML}(\mathbf{y}) = \text{sgn}(y_k)$. The relationship $V^{\varphi}(\mathbf{b}) \subseteq V^{\phi}(\mathbf{b})$ in (i) means that the fixed point region of detector $\varphi$ is inside that of $\phi$, which implies the relationship $\Psi^{\varphi}(\mathbf{y}) \subseteq \Psi^{\phi}(\mathbf{y})$ in (ii) that a fixed point of $\varphi$ is also a fixed point of $\phi$. $\phi$ performs worse than $\varphi$ only in the region $V^{\phi}(\mathbf{b}) \backslash V^{\varphi}(\mathbf{b})$.

Because their decision region is not inside a fixed point region of the LAS detector, none of linear detectors, SIC, PIC, DDF, and MMSE-DF [1] can be a fixed point of any LAS detector with probability one. By Theorem 1, any LAS detector can reduce their error probabilities and a WSLAS detector can reduce to local minima. All the LML detectors [37] with neighborhood size two, three, up to $K-1$, are also LML detectors with neighborhood size one. Their decision regions enclose $V^{GML}(\mathbf{b})$ and are enclosed by $V^{WSLAS}(\mathbf{b})$. Hence, a LAS detector does not change the initial LML detectors.

In Fig. 1 (a), the slopped line across the origin is the boundary of the GML decision regions for (−1,+1) and (+1,−1). The LML regions for (+1,+1) and (−1,−1) are identical to the GML regions. However, in the shaded region the LML regions for (+1,−1) and (−1,+1) are overlapped where both (+1,−1) and (−1,+1) are LML points. Only in this region, may an LML detector perform worse than the GML detector. Given any **y** in the overlapped region, an LML detector can take either (+1,−1) or (−1,+1) while the GML detector always chooses the maximum likelihood one. There are infinitely many LML detectors due to the infinitely many **y** each having two choices on (+1,−1) and (−1,+1). As an LML detector, a WSLAS detector converges to either (+1,−1) or (−1,+1) depending on the sequence of $L(n)$ and $\mathbf{b}(0)$. The BER upper bound to be obtained in the next section is applicable to all the LML/WSLAS detectors and thus is for the worst LML/WLAS detector that always chooses the error vector in the overlapped region. In Fig. 1 (b), the lightly shaded regions have two fixed points and the deeply shaded regions have three fixed points. The PLAS detector with the higher threshold performs worse than the SLAS detector.

## III. BIT ERROR PERFORMANCE

*A. The upper bound of BER*

Let $F_k$ be the set of indecomposable error vectors affecting user $k$ and $w(\varepsilon)$ be the weight of error vector $\varepsilon$. The main result of the paper is the following theorem which is proved in the next section.

*Theorem 2*: Given any initial detector $\mathbf{b}(0)$, the BER of the $k$th user for the LAS detector associated with a sequence of $L(n)$, $n \geq 0$ is upper bounded by

$$P_k^{\text{LAS}}(\sigma) \leq \sum_{\varepsilon \in F_k} 2^{-w(\varepsilon)} Q\left(\frac{\varepsilon^T (2\mathbf{H} - \mathbf{T})\varepsilon}{\sigma\sqrt{\varepsilon^T \mathbf{H} \varepsilon}}\right) \tag{11}$$

where $\mathbf{T} = \text{diag}(t_1^*, ..., t_K^*)$. □

The upper bound is applicable to all initial detectors since it is obtained by bounding the fixed point region that is the union of all the regions where a bit vector is a fixed point. Confirming the observation on the fixed point region, decrease of the thresholds can decrease the upper bound. Therefore, three changes can result in decrease of the upper bound: (i) decrease of the total number of update candidates; (ii) decrease of the absolute values of the crosscorrelations between the update candidates; and (iii) decrease of the signal powers of the update candidates.

By letting $t_i^* = A_i^2$ in (11), the BER upper bound for all the WSLAS/LML detectors is then obtained

$$P_k^{\text{LML}}(\sigma) \leq \sum_{\varepsilon \in F_k} 2^{-w(\varepsilon)} Q\left(\frac{\varepsilon^T (2\mathbf{H} - \mathbf{A}^2)\varepsilon}{\sigma\sqrt{\varepsilon^T \mathbf{H} \varepsilon}}\right). \tag{12}$$

It is clear that the WSLAS detectors achieve the least upper bound in the family. All the LML detectors with any neighborhood size [37] are also LML detectors with neighborhood size one. Hence, the upper bound (12) is applicable to all the LML detectors with any neighborhood size.

Given a CDMA channel ($\mathbf{S}$, $\mathbf{A}$), one can obtain the BER upper bounds (11) and (12) by searching the set of indecomposable error vectors in $E$ that has $|E| = 3^K - 1$ vectors. In order to reduce the searching time, the following lemma can be applied. For an index set $D \subseteq \{1, ..., K\}$, by $\psi(D) = \{\varepsilon \in E \mid I(\varepsilon) = D\}$ we define a set of error vectors $\varepsilon$ that have the same index set of nonzero elements $I(\varepsilon) = D$. There are a total of $|\psi(D)| = 2^{|D|}$ error vectors in $\psi(D)$. Let $F$ be the set of all indecomposable error vectors.

*Lemma 1*: For any nonempty $D \subseteq \{1, ..., K\}$, $\psi(D)$ has either none or two antipodal indecomposable error vectors; moreover, the total number of indecomposable error vectors in $E$ is upper bounded by $|F| \leq 2(2^K - 1)$. □

When searching for the indecomposable error vectors with given $D$, once an indecomposable error vector is obtained in $\psi(D)$, the other antipodal one is also obtained and then the search in $\psi(D)$ shall be terminated. The search time is then reduced from the order of $O(3^K)$ to $O(2^K)$.

*B. Comparison with the GML detector*

The BER upper bound (11) of the LAS detector is comparable with the upper bound of the GML detector obtained by Verdú [1]

$$P_k^{\text{GML}}(\sigma) \leq \sum_{\varepsilon \in F_k} 2^{-w(\varepsilon)} Q\left(\frac{\sqrt{\varepsilon^T \mathbf{H} \varepsilon}}{\sigma}\right). \tag{13}$$

Each pair of transmitted **b** and erroneous **b**$^*$ is associated with an error vector $\varepsilon = \frac{1}{2}(\mathbf{b} - \mathbf{b}^*)$. To obtain the upper bound (13), the **r** space is divided by the hyperplane $l^{\text{GML}}(\varepsilon)$ that separates the transmitted signal **SAb** and the error signal **SAb**$^*$ optimally in terms of GML. $l^{\text{GML}}(\varepsilon)$ has the equal distance to **SAb** and **SAb**$^*$ and contains the boundary of the GML decision regions of these two signals. The $Q$ function with $\varepsilon$ in the upper bound is the probability that the received signal **r** is located in the half space containing the GML decision region of **SAb**$^*$. The distance from **SAb** to this hyperplane equals

$$d^{\text{GML}}(\varepsilon) = \sqrt{\varepsilon^T \mathbf{H} \varepsilon}. \tag{14}$$

In comparison, to obtain the upper bound (11) for the LAS detector, the **r** space is divided by the hyperplane $l^{\text{LAS}}(\varepsilon)$ that passes through the vertex of the fixed point region of **SAb**$^*$ and is parallel to the optimal hyperplane $l^{\text{GML}}(\varepsilon)$. The corresponding $Q$ function in (11) is the probability that the received signal **r** is located in the half space containing the fixed point region of **SAb**$^*$. The distance from **SAb** to $l^{\text{LAS}}(\varepsilon)$ is equal to

$$d^{\text{LAS}}(\varepsilon) = \frac{\varepsilon^T (2\mathbf{H} - \mathbf{T})\varepsilon}{\sqrt{\varepsilon^T \mathbf{H} \varepsilon}}. \tag{15}$$

The difference between these two distances of the LAS and the GML detectors equals

$$\Delta d^{\text{LAS}}(\varepsilon) = d^{\text{LAS}}(\varepsilon) - d^{\text{GML}}(\varepsilon) = \frac{\varepsilon^T (\mathbf{H} - \mathbf{T})\varepsilon}{\sqrt{\varepsilon^T \mathbf{H} \varepsilon}} \tag{16}$$

which yields the difference of the upper bounds between them. Similarly, the distance difference between the LML and the GML detectors equals

$$\Delta d^{\text{LML}}(\varepsilon) = d^{\text{LML}}(\varepsilon) - d^{\text{GML}}(\varepsilon) = \frac{\varepsilon^T (\mathbf{H} - \mathbf{A}^2)\varepsilon}{\sqrt{\varepsilon^T \mathbf{H} \varepsilon}}. \tag{17}$$

If $w(\varepsilon) = 1$, which means $\varepsilon_k \neq 0$ and $\varepsilon_j = 0$ for $j \neq k$, then $\Delta d^{\text{LML}}(\varepsilon) = 0$. This is true since an LML detector achieves the maximum likelihood in a neighborhood of size one and therefore the boundary between the LML point regions for the two vectors different by one bit is the same as that of the GML decision regions. If $w(\varepsilon) = 2$, say $\varepsilon$ with $\varepsilon_k = 1$, $\varepsilon_j = -\text{sgn}(R_{kj})$ for $j \neq k$, and $\varepsilon_i = 0$ for $i \neq k$ and $i \neq j$,

$$\Delta d^{\text{LML}}(\varepsilon) = \frac{-2|R_{kj}|A_k A_j}{\sqrt{A_k^2 + A_j^2 - 2|R_{kj}|A_k A_j}},$$

which is less than zero as expected. In fact, we have the following lemma.

*Lemma 2*: For any $\varepsilon \in F_k$, $\varepsilon^T(\mathbf{H} - \mathbf{A}^2)\varepsilon \leq 0$ with equality iff $w(\varepsilon) = 1$. □

The lemma implies that if $\boldsymbol{\varepsilon}$ is *indecomposable*, then $\boldsymbol{\varepsilon}^T(\mathbf{H}-\mathbf{T})\boldsymbol{\varepsilon} \leq \boldsymbol{\varepsilon}^T(\mathbf{H}-\mathbf{A}^2)\boldsymbol{\varepsilon} \leq 0$ with the second equality if and only if $w(\boldsymbol{\varepsilon}) = 1$. Since only the indecomposable error vectors affect the upper bounds, the $Q$ function evaluated with any $\boldsymbol{\varepsilon}$ in the upper bound (11) is not less than the corresponding term in (13). Hence, the LAS upper bound is not less than the GML upper bound. However, it shall be pointed out that $\Delta d^{\text{LAS}}(\boldsymbol{\varepsilon}) > 0$ is possible for an arbitrary $\boldsymbol{\varepsilon}$. For example, in the two-user channel of Fig. 1 (a), $\Delta d^{\text{LAS}}(\boldsymbol{\varepsilon}) > 0$ when $(-1,-1)$ is transmitted and the detected vector is $(+1,+1)$. In this case, the error vector $\boldsymbol{\varepsilon} = (-1,-1)$ is decomposable.

*C. The lower bound of AME*

The asymptotical multiuser efficiency (AME) measures in the high SNR regime the efficiency of the signal power usage by a detector compared with the MF in the single-user channel [1]. By (11), the following corollary is obtained.

*Corollary 1*: The AME of the LAS detector is lower bounded by

$$\eta_k^{\text{LAS}} \geq \min_{\boldsymbol{\varepsilon} \in F_k}{}^2 \left\{ \frac{[\boldsymbol{\varepsilon}^T(2\mathbf{H}-\mathbf{T})\boldsymbol{\varepsilon}]^+}{A_k \sqrt{\boldsymbol{\varepsilon}^T \mathbf{H} \boldsymbol{\varepsilon}}} \right\}; \tag{18}$$

in particular, the AME for all the WSLAS and LML detectors is lower bounded by

$$\eta_k^{\text{LML}} \geq \min_{\boldsymbol{\varepsilon} \in F_k}{}^2 \left\{ \frac{[\boldsymbol{\varepsilon}^T(2\mathbf{H}-\mathbf{A}^2)\boldsymbol{\varepsilon}]^+}{A_k \sqrt{\boldsymbol{\varepsilon}^T \mathbf{H} \boldsymbol{\varepsilon}}} \right\} \tag{19}$$

where $[z]^+ = \max\{0, z\}$. □

The lower bound (19) is also applicable to all the LML detectors with neighborhood size greater than one [37].

*D. Achievability of unit AME*

Unit AME is particularly interesting in CDMA. In high SNR regime, BER is dominated by the minimum distance from the transmitted signal to the decision regions of error signals. If AME for user $k$ is unit, the minimum distance is determined by the single error and then user $k$'s BER achieves the single user bound $Q(A_k/\sigma)$ as if there was no interferer. However, none of the well-known suboptimum detectors is known to achieve unit AME except in some trivial two-user channels and the orthogonal $K$-user channel [1].

It follows from (13) that the GML detector achieves unit AME in the channels such that

$$\sqrt{\boldsymbol{\varepsilon}^T \mathbf{H} \boldsymbol{\varepsilon}} \geq A_k, \forall \boldsymbol{\varepsilon} \in F_k. \tag{20}$$

In these channels, standing at the transmitted signal, one sees that the nearest error signals with the $k$th bit being erroneous are those that only the $k$th bit is erroneous and all other bits are correct. If this is true for all $k$, then one sees that the nearest error signals are those that have only one error bit. The BER of the GML detector in the high SNR region is then dominated by the single-error signals and approaches the single-user bound asymptotically. The result can be regardless of $K$. The similar result can be obtained for the LAS detector.

*Corollary 2*: If for each $\boldsymbol{\varepsilon} \in F_k$ with each $w(\boldsymbol{\varepsilon}) = m \leq K$ where $K$ can be finite or tend to infinity

$$\frac{\varepsilon^T (2\mathbf{H} - \mathbf{T})\varepsilon}{\sqrt{\varepsilon^T \mathbf{H}\varepsilon}} \geq A_k , \qquad (21)$$

then the LAS detector achieves unit AME for user $k$. Furthermore, if (21) holds for all users, then the LAS detector achieve unit AME for all users; in particular, these are true for all LML detectors with $\mathbf{T}$ in (21) replaced by $\mathbf{A}^2$. □

For the channels characterized by (21), standing at the transmitted signal, one sees that the nearest fixed point region of error signals whose $k$th bit is erroneous are of those error vectors of which only the $k$th bit is erroneous and all other bits are correct. If this is true for all $k$, one sees that the nearest fixed point regions of error signals are of those error vectors having only one error bit. The BER of the LAS detector in the high SNR region is then dominated by these single-error signals and approaches the single-user bound asymptotically. The result can be regardless of $K$. It is clear that condition (21) implies (20). In fact, in the channels characterized by (21) all the LML detectors with any neighborhood size [37] achieve unit AME.

There exist many CDMA channels that satisfy (21) regardless of $K$. For example, consider the channel where the users have the same power $A_i = A$ and the same crosscorrelation $R_{ij} = \rho > 0$, $\forall i \neq j$. The set of indecomposable error vectors for user $k$ consists of the error vectors $(1, 0, 0, \ldots, 0)$, $(1, -1, 0, \ldots, 0)$, $\ldots$, $(1, 0, 0, \ldots, -1)$ and their antipodal images [1]. The BER upper bounds of both the GML and the LAS detectors depend only on the error vectors that have one or two errors. The GML detector for all users attains the AME $\eta^{\text{GML}} = \min\{1, 2(1-\rho)\}$, which equals one if $\rho \leq \frac{1}{2}$ regardless of $K$.

For a GPLAS detector that updates $M$ bits in each step, the threshold equals $t_k = A^2 + (M-1)\rho A^2$. Each of the error vectors produces the same $Q$-function term in the upper bound. It follows from (18) that the AME for all users is lower bounded by

$$\eta^{\text{LAS}} \geq \min^2 \left\{ 1, \frac{2[1-(M+1)\rho]^+}{\sqrt{2(1-\rho)}} \right\}. \qquad (22)$$

Letting $M = 1$ yields a lower bound for all the LML detectors

$$\eta^{\text{LML}} \geq \min^2 \left\{ 1, \frac{2[1-2\rho]^+}{\sqrt{2(1-\rho)}} \right\}. \qquad (23)$$

Then the following proposition is obtained.

*Proposition 2*: In a channel that has equal user power and identical positive crosscorrelation $\rho$, the GPLAS detector with the same group size of $M$ achieves unit AME if

$$\rho \leq \rho_M = \frac{4M + 3 - \sqrt{8M^2 + 8M + 1}}{4(M+1)^2} ; \qquad (24)$$

in particular, if $\rho \leq \rho_1 = (7 - \sqrt{17})/16$, all the LML detectors achieve unit AME. □

It is interesting to note that the GPLAS and the LML detectors achieve unit AME in the conditions of Proposition 2 regardless of $K$. This is due to that given any $K$, the upper bounds of BER for the LAS and LML detectors depend only on the error vectors that have one or two errors. The signals with three or more errors are farther than those with one and two errors. Alike the GML detector, the LAS detector exploits the LML characteristic of this channel.

In the channel of Fig. 2, the AME of the MF is obtained by the formula $\eta^{MF} = \max^2\{0, 1-(K-1)\rho\}$ [1]. The decorrelator and the MMSE have the same AME that is equal to the reciprocal of the $k$th diagonal element of $\mathbf{R}^{-1}$ for the $k$th user, $\eta^{DEC/MMSE} = [1 + (K-2)\rho - (K-1)\rho^2]/[1 + (K-2)\rho]$, which is equal to $1 - \rho$ for sufficiently large $K$. The MF and the decorrelator/MMSE are unable to achieve unit AME except in the trivial case of $\rho = 0$. In contrast, the GPLAS and the LML detectors can achieve unit AME regardless of $K$. The AME's of all other LML detectors with neighborhood size greater than one and less than $K$ in [37] must be located in the region between the AME lower bound of the LML detectors and the AME of the GML detector in the figure.

## IV. PROOF OF THE UPPER BOUND

Let $A(\mathbf{b})$ be the set of admissible error vectors of $\mathbf{b}$, $A_k(\mathbf{b})$ be the set of admissible error vectors of $\mathbf{b}$ affecting user $k$, and $E_k$ be the set of error vectors affecting user $k$.

The proof can be visualized geometrically by Fig. 3 shows the LML region $V(\mathbf{b}-2\boldsymbol{\varepsilon})$ in the $\mathbf{y}$ space for a two-user channel. The $\mathbf{y}$ space is transformed from the $\mathbf{r}$ space by $\mathbf{S}^T$. In what follows, the prime $'$ after a notation means the corresponding hyperplane or region in the $\mathbf{r}$ space. For example, $V'(\mathbf{b})$ denotes the fixed point region of $\mathbf{b}$ in the $\mathbf{r}$ space, which is corresponding to the fixed point region $V(\mathbf{b})$ in the $\mathbf{y}$ space.

Consider in the $\mathbf{r}$ space the transmitted signal $\mathbf{SAb}$ and the error signal $\mathbf{SA}(\mathbf{b}-2\boldsymbol{\varepsilon})$ with the error vector $\boldsymbol{\varepsilon}$. The midpoint of the line segment from $\mathbf{SA}(\mathbf{b}-2\boldsymbol{\varepsilon})$ to $\mathbf{SAb}$ is $[\mathbf{SA}(\mathbf{b}-2\boldsymbol{\varepsilon}) + \mathbf{SAb}]/2 = \mathbf{SA}(\mathbf{b}-\boldsymbol{\varepsilon})$. The vector from $\mathbf{SA}(\mathbf{b}-2\boldsymbol{\varepsilon})$ to $\mathbf{SAb}$ is $\mathbf{SAb} - \mathbf{SA}(\mathbf{b}-2\boldsymbol{\varepsilon}) = 2\mathbf{SA}\boldsymbol{\varepsilon}$. The hyperplane $c'$ that passes through the boundary between the GML decision regions of signals $\mathbf{SAb}$ and $\mathbf{SA}(\mathbf{b}-2\boldsymbol{\varepsilon})$ is $\mathbf{r}^T(2\mathbf{SA}\boldsymbol{\varepsilon}) = [\mathbf{SA}(\mathbf{b}-\boldsymbol{\varepsilon})]^T(2\mathbf{SA}\boldsymbol{\varepsilon})$ or $\mathbf{y}^T\mathbf{A}\boldsymbol{\varepsilon} = [\mathbf{SA}(\mathbf{b}-\boldsymbol{\varepsilon})]^T\mathbf{SA}\boldsymbol{\varepsilon}$. By (9), the fixed point region of the bit vector $(\mathbf{b}-2\boldsymbol{\varepsilon})$ is

$$V(\mathbf{b} - 2\boldsymbol{\varepsilon}) = \{\mathbf{y} \in \Re^K \mid (\mathbf{b} - 2\boldsymbol{\varepsilon}) \otimes [\mathbf{Ay} - \mathbf{H}(\mathbf{b} - 2\boldsymbol{\varepsilon})] \geq -\mathbf{t}^*\} \tag{25}$$

and its vertex is at $\mathbf{y} = (\mathbf{RA} - \mathbf{A}^{-1}\mathbf{T})(\mathbf{b} - 2\boldsymbol{\varepsilon})$ or

$$\mathbf{S}^T\mathbf{v} = (\mathbf{RA} - \mathbf{A}^{-1}\mathbf{T})(\mathbf{b} - 2\boldsymbol{\varepsilon}) \tag{26}$$

where the vertex in the $\mathbf{r}$ space is denoted by $\mathbf{v}$. The hyperplane $d'$ that is parallel to hyperplane $c'$ and passes through the vertex is $\mathbf{r}^T\mathbf{SA}\boldsymbol{\varepsilon} = \mathbf{v}^T\mathbf{SA}\boldsymbol{\varepsilon}$ or

$$\mathbf{y}^T\mathbf{A}\boldsymbol{\varepsilon} = \mathbf{v}^T\mathbf{SA}\boldsymbol{\varepsilon} . \tag{27}$$

The half space $\Omega'(\mathbf{b}-2\boldsymbol{\varepsilon})$ that is divided out by the hyperplane $d'$ and contains the error signal $\mathbf{SA}(\mathbf{b}-2\boldsymbol{\varepsilon})$ is the collection of points $\mathbf{r}$ that satisfy $(\mathbf{r} - \mathbf{v})^T(2\mathbf{SA}\boldsymbol{\varepsilon}) \leq 0$ or $\mathbf{r}^T\mathbf{SA}\boldsymbol{\varepsilon} \leq \mathbf{v}^T\mathbf{SA}\boldsymbol{\varepsilon}$. It follows from (26) that

$$\Omega(\mathbf{b} - 2\boldsymbol{\varepsilon}) = \{\mathbf{y} \in \Re^K \mid \mathbf{y}^T\mathbf{A}\boldsymbol{\varepsilon} \leq (\mathbf{b} - 2\boldsymbol{\varepsilon})^T(\mathbf{H} - \mathbf{T})\boldsymbol{\varepsilon}\} . \tag{28}$$

The following lemma indicates that the probability for $\mathbf{y} \in \Omega(\mathbf{b}-2\boldsymbol{\varepsilon})$ is an upper bound on the probability for $\mathbf{y} \in V(\mathbf{b}-2\boldsymbol{\varepsilon})$.

*Lemma 3*: $V(\mathbf{b}-2\boldsymbol{\varepsilon}) \subseteq \Omega(\mathbf{b}-2\boldsymbol{\varepsilon})$ for any $\mathbf{b} \in \{-1,1\}^K$ and $\boldsymbol{\varepsilon} \in A(\mathbf{b})$. □

Then the BER of the LAS detector with any initial vector and any sequence of $L(n)$ is upper bounded by

$$P_k^{\text{LAS}}(\sigma) \leq \sum_{\boldsymbol{\varepsilon} \in E_k} \Pr[\boldsymbol{\varepsilon} \in A_k(\mathbf{b}); \mathbf{y} \in V(\mathbf{b}-2\boldsymbol{\varepsilon})] \leq \sum_{\boldsymbol{\varepsilon} \in E_k} \Pr[\boldsymbol{\varepsilon} \in A_k(\mathbf{b}); \mathbf{y} \in \Omega(\mathbf{b}-2\boldsymbol{\varepsilon})] \quad (29)$$

where the first inequality is due to that $V(\mathbf{b}-2\boldsymbol{\varepsilon})$ may be overlapped with the fixed point region of the transmitted $\mathbf{b}$ and the second follows from Lemma 3. Given $\boldsymbol{\varepsilon} \in E_k$, there are a total of $2^{w(\boldsymbol{\varepsilon})}$ bit vectors that satisfy $\boldsymbol{\varepsilon} \in A_k(\mathbf{b})$ and each of which occurs with equal probability $\Pr[\boldsymbol{\varepsilon} \in A_k(\mathbf{b})] = 2^{-w(\boldsymbol{\varepsilon})}$.

The event $\mathbf{y} \in \Omega(\mathbf{b} - 2\boldsymbol{\varepsilon})$ implies $\mathbf{y}^T \mathbf{A}\boldsymbol{\varepsilon} \leq (\mathbf{b} - 2\boldsymbol{\varepsilon})^T (\mathbf{H} - \mathbf{T})\boldsymbol{\varepsilon}$. Replacing $\mathbf{y} = \mathbf{RAb} + \mathbf{z}$ yields $\mathbf{z}^T \mathbf{A}\boldsymbol{\varepsilon} \leq -\mathbf{b}^T \mathbf{H}\boldsymbol{\varepsilon} + (\mathbf{b} - 2\boldsymbol{\varepsilon})^T(\mathbf{H} - \mathbf{T})\boldsymbol{\varepsilon} = -2\boldsymbol{\varepsilon}^T \mathbf{H}\boldsymbol{\varepsilon} - (\mathbf{b} - 2\boldsymbol{\varepsilon})^T \mathbf{T}\boldsymbol{\varepsilon} = -\boldsymbol{\varepsilon}^T(2\mathbf{H} - \mathbf{T})\boldsymbol{\varepsilon}$ where the last equality is due to $(\mathbf{b} - 2\boldsymbol{\varepsilon})^T \mathbf{T}\boldsymbol{\varepsilon} = -\boldsymbol{\varepsilon}^T \mathbf{T}\boldsymbol{\varepsilon}$ because if $\varepsilon_i \neq 0$, then $b_i - 2\varepsilon_i = -b_i = -\varepsilon_i$. Since $\mathbf{z}^T \mathbf{A}\boldsymbol{\varepsilon}$ is a Gaussian variable with mean zero and variance $\sigma^2 \boldsymbol{\varepsilon}^T \mathbf{H}\boldsymbol{\varepsilon}$,

$$\Pr[\mathbf{y} \in \Omega(\mathbf{b} - 2\boldsymbol{\varepsilon})] \leq Q\left(\frac{\boldsymbol{\varepsilon}^T (2\mathbf{H} - \mathbf{T})\boldsymbol{\varepsilon}}{\sigma\sqrt{\boldsymbol{\varepsilon}^T \mathbf{H}\boldsymbol{\varepsilon}}}\right). \quad (30)$$

In fact, the upper bound (30) can be also obtained in terms of the distance from $\mathbf{SAb}$ to hyperplane $d'$, which is equal to $B = (\mathbf{SAb} - \mathbf{v})^T \mathbf{SA}\boldsymbol{\varepsilon}/\|\mathbf{SA}\boldsymbol{\varepsilon}\| = [\mathbf{RAb} - (\mathbf{RA} - \mathbf{A}^{-1}\mathbf{T})(\mathbf{b} - 2\boldsymbol{\varepsilon})]^T \mathbf{A}\boldsymbol{\varepsilon}/\|\mathbf{SA}\boldsymbol{\varepsilon}\| = [2\mathbf{RA}\boldsymbol{\varepsilon} + \mathbf{A}^{-1}\mathbf{T}(\mathbf{b} - 2\boldsymbol{\varepsilon})]^T \mathbf{A}\boldsymbol{\varepsilon}/\|\mathbf{SA}\boldsymbol{\varepsilon}\|$ $= [2\boldsymbol{\varepsilon}^T \mathbf{H}\boldsymbol{\varepsilon} + (\mathbf{b} - 2\boldsymbol{\varepsilon})^T \mathbf{T}\boldsymbol{\varepsilon}]/\|\mathbf{SA}\boldsymbol{\varepsilon}\| = \boldsymbol{\varepsilon}^T(2\mathbf{H} - \mathbf{T})\boldsymbol{\varepsilon}/(\boldsymbol{\varepsilon}^T \mathbf{H}\boldsymbol{\varepsilon})^{\frac{1}{2}}$. Since the noise vector in the $\mathbf{r}$ space is white Gaussian with variance $\sigma^2$ for each component, we obtain (30) by $Q(B/\sigma)$. If the distance is negative, then $\mathbf{S}^T \mathbf{SAb} \in \Omega(\mathbf{b}-2\boldsymbol{\varepsilon})$ and (30) with $Q(B/\sigma) > \frac{1}{2}$ is invalid. However, in most cases the upper bound is tight and valid.

The events $\boldsymbol{\varepsilon} \in A_k(\mathbf{b})$ and $\mathbf{y} \in \Omega(\mathbf{b} - 2\boldsymbol{\varepsilon})$ are mutually independent since they depend only on $\mathbf{b}$ and $\mathbf{z}$, respectively. Hence,

$$P_k^{LAS}(\sigma) \leq \sum_{\boldsymbol{\varepsilon} \in E_k} 2^{-w(\boldsymbol{\varepsilon})} Q\left(\frac{\boldsymbol{\varepsilon}^T (2\mathbf{H} - \mathbf{T})\boldsymbol{\varepsilon}}{\sigma\sqrt{\boldsymbol{\varepsilon}^T \mathbf{H}\boldsymbol{\varepsilon}}}\right). \quad (31)$$

The notion of indecomposable errors developed by Verdú to tighten the GML BER upper bound [1] can be also applied to tighten the LAS BER upper bound. The following lemma indicates that the terms produced by the decomposable errors in (31) are redundant. Hence, $E_k$ in (29) can be replaced by $F_k$ and this finishes the proof of Theorem 2.

*Lemma 4*: For every $\mathbf{b} \in \{-1,1\}^K$ and $\boldsymbol{\varepsilon} \in A_k(\mathbf{b})$, there is $\boldsymbol{\varepsilon}^* \in F_k$ s. t. $V(\mathbf{b} - 2\boldsymbol{\varepsilon}) \subseteq \Omega(\mathbf{b} - 2\boldsymbol{\varepsilon}^*)$. □

## V. CONCLUSIONS

The family of the LAS detectors has many good common characteristics, mostly brought out by the characteristic that the sequence of bit vectors is monotonically increasing. This implies the convergence to a fixed point in a finite number of steps as well as the monotonic decrease of error probability. Thus, a LAS detector always reduces the error probability of an initial detector which is not a fixed point with probability one. The WSLAS detector reduces the initial error probability to a local minimum. Numerous simulation results, part of which is reported in [39], show that the per-user complexity of any LAS detector with the initial MF is less than 0.5 times the user number.

The LAS BER upper bound is comparable with the GML BER upper bound obtained by Verdú. The concept of indecomposable error vectors developed by Verdú to tighten the GML upper bound is also applicable to tighten the LAS upper bound. As a special instance, the LAS upper bound becomes an upper bound for all the LML detectors. The upper bound decreases as the number of users updated simultaneously decreases. The performance gap between a LAS detector and the GML detector is clearly shown due to the shorter distance of a LAS detector from a transmitted signal to the fixed point region of an error signal. The LAS AME lower bound indicates that there exist nontrivial CDMA channels where a LAS detector achieves unit AME regardless of user number. Further analysis in [39] indicates that the LAS detector can achieve unit AME almost surely in large random spreading CDMA when channel load is less than ½ − 1/(4ln2) bits/s/Hz. These in turn suggest that the AME lower bound be useful in design of spectral and power efficient CDMA systems where the LAS detector is employed.

## APPENDIX

The following lemma is useful in analysis with its proof left for reader.

*Lemma 5*: Let $P_e(\mathbf{b}^\varphi)$ and $P_e(\mathbf{b}^\phi)$ be the error probabilities of two detectors $\mathbf{b}^\varphi$ and $\mathbf{b}^\phi$, respectively. If $f(\mathbf{y} \mid \mathbf{b}^\varphi) \geq f(\mathbf{y} \mid \mathbf{b}^\phi)$ for all $\mathbf{y} \in \mathfrak{R}^K$, then $P_e(\mathbf{b}^\varphi) \leq P_e(\mathbf{b}^\phi)$ with equality iff $\Pr[f(\mathbf{y} \mid \mathbf{b}^\phi) = f(\mathbf{y} \mid \mathbf{b}^\varphi)] = 1$. □

*Proof of Proposition 1*:

Denote by $g_k(\mathbf{b})$ the $k$th component of the likelihood gradient at the fixed point $\mathbf{b}$. For each $\mathbf{y} \in \mathfrak{R}^K$, it follows from (7) that for any $\mathbf{b}(0) \in \Omega(\mathbf{y},\mathbf{b})$ the fixed point must satisfy $b_k g_k(\mathbf{b}) \geq \max_{n \geq n_k^*[\mathbf{b}(0)], k \in L(n)} -t_k(n) = -\min_{n \geq n_k^*[\mathbf{b}(0)], k \in L(n)} t_k(n)$. To include all possible initial vectors,

$$b_k g_k(\mathbf{b}) \geq \min_{\mathbf{b}(0) \in \Omega(\mathbf{y},\mathbf{b})} \{-\min_{n \geq n_k^*[\mathbf{b}(0)], k \in L(n)} t_k(n)\} = -t_k^*, \forall k. \tag{32}$$

Since $b_k g_k(\mathbf{b}) = b_k \left(-\sum_{j=1}^K H_{kj} b_j + A_k y_k\right)$ from (3), (32) yields the fixed-point region. □

*Proof of Lemma 1*:

Obviously, it is possible that $\psi(D)$ contains no indecomposable error vector.

Suppose there is an indecomposable error vector $\boldsymbol{\varepsilon} \in \psi(D) \cap F$. Then it is clear that $-\boldsymbol{\varepsilon} \in \psi(D) \cap F$. In what follows, it is shown that there is no other indecomposable error vector in $\psi(D)$. Since $\boldsymbol{\varepsilon}$ is indecomposable, $\boldsymbol{\varepsilon}_1^T \mathbf{H} \boldsymbol{\varepsilon}_2 < 0$ for any $\boldsymbol{\varepsilon}_1 \in E$ and $\boldsymbol{\varepsilon}_2 \in E$ such that (i) $\boldsymbol{\varepsilon} = \boldsymbol{\varepsilon}_1 + \boldsymbol{\varepsilon}_2$; (ii) if $\varepsilon_i = 0$, then $\varepsilon_{1i} = \varepsilon_{2i} = 0$. Suppose $\mathbf{a} \in \psi(D)$ is the third indecomposable error vector. Then $\mathbf{a}$ must have a number of nonzero elements, less than $|D|$, that are antipodal to the corresponding nonzero elements of $\boldsymbol{\varepsilon}$. Specifically, it can be written into $\mathbf{a} = \mathbf{a}_1 + \mathbf{a}_2$ where $\mathbf{a}_1, \mathbf{a}_2 \in E$, $a_{1i} = -\varepsilon_i$ for $i \in I(\mathbf{a}_1) \neq \varnothing$, $a_{2i} = \varepsilon_i$ for $i \in I(\mathbf{a}_2) \neq \varnothing$, and $I(\mathbf{a}_1) \cap I(\mathbf{a}_2) = \varnothing$. The indecomposability of $\mathbf{a}$ requires that $\mathbf{a}_1^T \mathbf{H} \mathbf{a}_2 < 0$. However, it implies that $-\mathbf{a}_1^T \mathbf{H} \mathbf{a}_2 > 0$, contradicting that $\boldsymbol{\varepsilon}$ is indecomposable. Hence, there is no third indecomposable error vector.

In $\{1, \ldots, K\}$, there are a total of $\binom{K}{|D|}$ sets that have the same number of elements $|D|$. Hence, the total number of indecomposable error vectors is upper bounded by $|F| \leq 2(2^K - 1)$. $\square$

*Proof of Lemma 2:*

It is obvious that if $w(\boldsymbol{\varepsilon}) = 1$, then $\boldsymbol{\varepsilon}^T(\mathbf{H}-\mathbf{A}^2)\boldsymbol{\varepsilon} = 0$.

For any $\boldsymbol{\varepsilon} \in F_k$ with $w(\boldsymbol{\varepsilon}) = M > 1$, consider $\boldsymbol{\varepsilon}', \boldsymbol{\varepsilon}'' \in E$ s. t. $\boldsymbol{\varepsilon} = \boldsymbol{\varepsilon}' + \boldsymbol{\varepsilon}''$, $w(\boldsymbol{\varepsilon}') = 1$, and $\varepsilon'_{(i)} = \varepsilon''_{(i)} = 0$ if $\varepsilon_{(i)} = 0$. Since $\boldsymbol{\varepsilon}$ is indecomposable, $\boldsymbol{\varepsilon}'^T \mathbf{H} \boldsymbol{\varepsilon}'' < 0$ for all such pairs of $\boldsymbol{\varepsilon}'$ and $\boldsymbol{\varepsilon}''$. Then we have

$$\boldsymbol{\varepsilon}^T(\mathbf{H}-\mathbf{A}^2)\boldsymbol{\varepsilon} = \sum_{i=1}^{M}\sum_{j=1, j\neq i}^{M} \varepsilon_{(i)} H_{(i)(j)} \varepsilon_{(j)} = \sum_{w(\boldsymbol{\varepsilon}')=1} \boldsymbol{\varepsilon}'^T \mathbf{H} \boldsymbol{\varepsilon}'' < 0$$

where $\varepsilon_{(i)}$ for $i = 1, \ldots, M$ are the $M$ nonzero elements of $\boldsymbol{\varepsilon}$. $\square$

*Proof of Lemma 3*:

If $\mathbf{y} \in V(\mathbf{b}-2\boldsymbol{\varepsilon})$, then by (9)

$$(b_i - 2\varepsilon_i)(A_i y_i - \sum_{j=1}^{K} H_{ij}(b_j - 2\varepsilon_j)) \geq -t_i^*. \tag{33}$$

If $\varepsilon_i \neq 0$, then $b_i - 2\varepsilon_i = -b_i = -\varepsilon_i$ and so $-\varepsilon_i(A_i y_i - \sum_{j=1}^{K} H_{ij}(b_j - 2\varepsilon_j)) \geq -t_i^*$ and thus

$$\varepsilon_i A_i y_i \leq \varepsilon_i \sum_{j=1}^{K} H_{ij}(b_j - 2\varepsilon_j) + t_i^* = \varepsilon_i \sum_{j=1}^{K}(H_{ij} - T_{ij})(b_j - 2\varepsilon_j). \tag{34}$$

Therefore,

$$\mathbf{y}^T \mathbf{A} \boldsymbol{\varepsilon} \leq \sum_{i=1}^{K} \varepsilon_i \sum_{j=1}^{K}(H_{ij} - T_{ij})(b_j - 2\varepsilon_j) = (\mathbf{b} - 2\boldsymbol{\varepsilon})^T (\mathbf{H} - \mathbf{T})\boldsymbol{\varepsilon}. \tag{35}$$

Hence, $\mathbf{y} \in \Omega(\mathbf{b}-2\boldsymbol{\varepsilon})$ and then $V(\mathbf{b}-2\boldsymbol{\varepsilon}) \subseteq \Omega(\mathbf{b}-2\boldsymbol{\varepsilon})$. $\square$

*Proof of Lemma 4:*

If $\boldsymbol{\varepsilon} \in A_k(\mathbf{b}) \cap F_k$, we take $\boldsymbol{\varepsilon} = \boldsymbol{\varepsilon}^*$. By Lemma 3, $V(\mathbf{b} - 2\boldsymbol{\varepsilon}) \subseteq \Omega(\mathbf{b} - 2\boldsymbol{\varepsilon}^*)$.

If $\boldsymbol{\varepsilon} \in A_k(\mathbf{b})$ but $\boldsymbol{\varepsilon} \notin F_k$, according to the proof of Proposition 4.2 in [1] $\boldsymbol{\varepsilon}$ can be decomposed into $\boldsymbol{\varepsilon} = \boldsymbol{\varepsilon}^* + \boldsymbol{\varepsilon}'$ s. t. $\boldsymbol{\varepsilon}^* \in F_k$ and $\boldsymbol{\varepsilon}^{*T}\mathbf{H}\boldsymbol{\varepsilon}' \geq 0$. For any $\mathbf{y} \in V(\mathbf{b} - 2\boldsymbol{\varepsilon})$, (34) holds for each $\varepsilon_i \neq 0$. If $\varepsilon_i^* \neq 0$, then $\varepsilon_i^* = \varepsilon_i$ and $\varepsilon_i' = 0$. Therefore, $\Omega(\mathbf{b} - 2\boldsymbol{\varepsilon}) = \{\mathbf{y} \in \Re^K \mid \mathbf{y}^T \mathbf{A}\boldsymbol{\varepsilon} \leq (\mathbf{b} - 2\boldsymbol{\varepsilon})^T (\mathbf{H} - \mathbf{T})\boldsymbol{\varepsilon}\}$ and

$$\mathbf{y}^T \mathbf{A} \boldsymbol{\varepsilon}^* \leq \sum_{\varepsilon_i^* \neq 0} \varepsilon_i^* \sum_{j=1}^{K}(H_{ij} - T_{ij})(b_j - 2\varepsilon_j) = (\mathbf{b} - 2\boldsymbol{\varepsilon})^T (\mathbf{H} - \mathbf{T})\boldsymbol{\varepsilon}^* = (\mathbf{b} - 2\boldsymbol{\varepsilon}^*)^T (\mathbf{H} - \mathbf{T})\boldsymbol{\varepsilon}^* - 2\boldsymbol{\varepsilon}'^T (\mathbf{H} - \mathbf{T})\boldsymbol{\varepsilon}^*$$

$$= (\mathbf{b} - 2\boldsymbol{\varepsilon}^*)^T (\mathbf{H} - \mathbf{T})\boldsymbol{\varepsilon}^* - 2\boldsymbol{\varepsilon}'^T \mathbf{H}\boldsymbol{\varepsilon}^* \leq (\mathbf{b} - 2\boldsymbol{\varepsilon}^*)^T (\mathbf{H} - \mathbf{T})\boldsymbol{\varepsilon}^*. \tag{36}$$

Hence, $\mathbf{y} \in \Omega(\mathbf{b}-2\boldsymbol{\varepsilon}^*)$ and then $V(\mathbf{b}-2\boldsymbol{\varepsilon}) \subseteq \Omega(\mathbf{b}-2\boldsymbol{\varepsilon}^*)$. $\square$

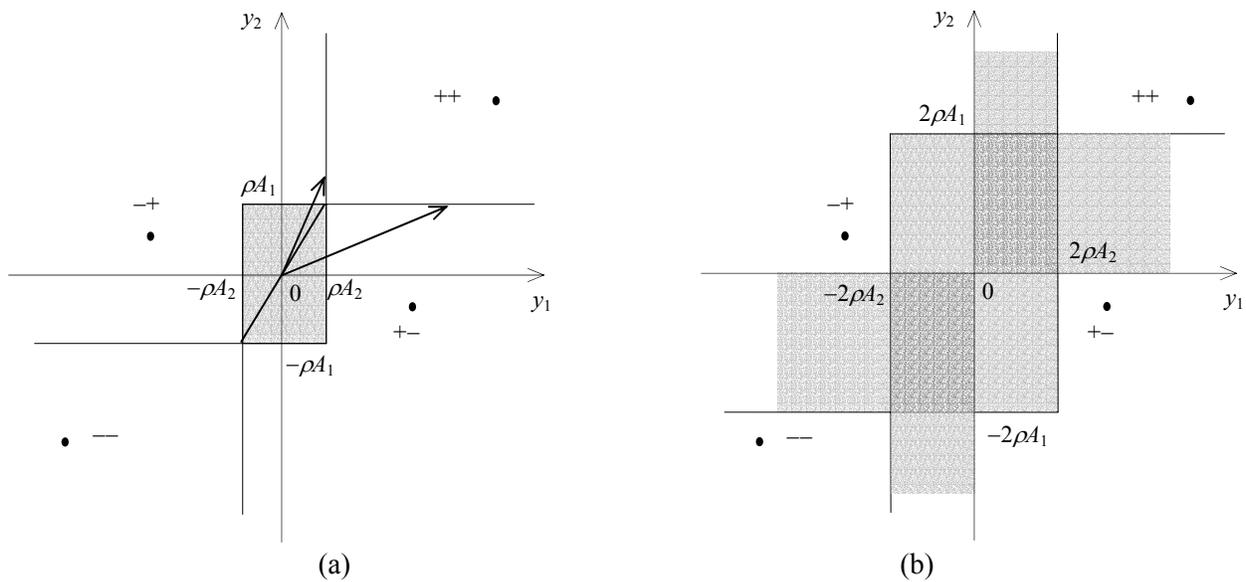

Fig. 1. The **y** space for the two-user channel with $\rho = 0.4$, $A_1 = 1$ and $A_2 = 0.6$. (a) The fixed point regions (or LML point regions) of the WSLAS detectors. (b) The fixed point regions of the PLAS detector.

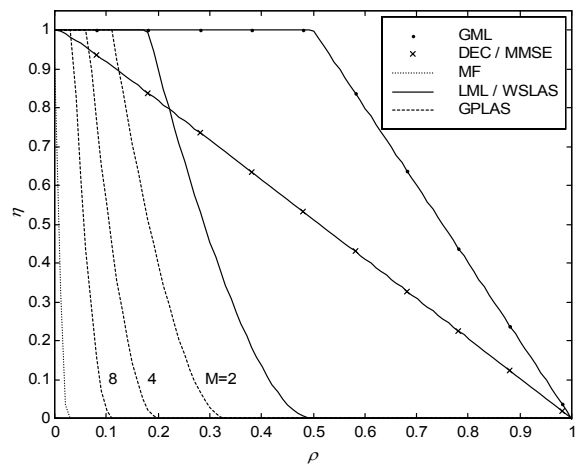

Fig. 2. The AME lower bounds of the GPLAS and LML detectors with equal power and identical crosscorrelations. The user number for the decorrelator/MMSE and MF is $K = 40$.

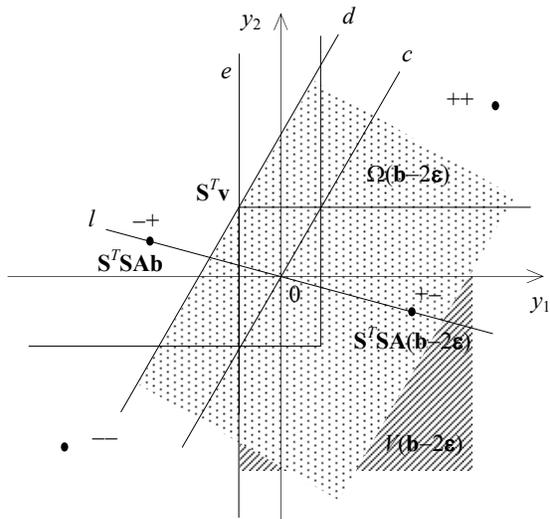

Fig. 3. The **y** space for a two-user channel. The probability for $\mathbf{y} \in \Omega(\mathbf{b}-2\boldsymbol{\varepsilon})$ (dotted region) is used as an upper bound for the probability for $\mathbf{y} \in V(\mathbf{b}-2\boldsymbol{\varepsilon})$ (stripped region) when **b** is transmitted.